# Development of an elementary climate model: two-layer cellular case

Laura E. Schmidt,[a,c] H. L. Helfer,[a,b] and Robert S. Knox[a,b,]*

[a]Department of Physics and Astronomy, University of Rochester, Rochester, NY 14627–0171
[b]Laboratory for Laser Energetics, University of Rochester, Rochester, NY 14623–1299
[c]Present address: Department of Physics, University of Chicago, Chicago, IL 60637–1434
*Corresponding author. Email `rsk@pas.rochester.edu`

## Abstract

A qualitative understanding of the greenhouse effect has long been available through models based on globally- and time-averaged quantities. We examine here a simple 864-cell climatological model that emphasizes vertical radiative energy transport within each cell. It reproduces yearly average temperatures obtained earlier from one of these global models and predicts a locally distributed non-radiative flux when observed temperatures are employed as input data. Vertical and lateral transport of latent heat do not appear explicitly in this model. They are apparently handled well by one non-radiative flux variable, $S_{NR}$, which shows a strong latitude dependence. Only the Sahara desert and Saudi Arabian regions appear to be complex. For those interested in climatology and construction of climate models, our model provides constraints upon specifying averaged non-radiative energy transport. The model is a useful stepping-stone for learning about radiative energy transfer into and out of Earth's atmosphere and for representing the results of more sophisticated models.

## I. INTRODUCTION

The recent gradual increase in Earth's average temperature [1] has generated interest in both public and scientific circles. The purpose of this paper is to construct a very simple model capable of representing results similar to those found by more sophisticated calculations while allowing physicists not in the field to focus on understanding the basic energetics in climatology. Introductory atmosphere and climatology texts such as Goody and Walker [2] and Hartmann [3] and the more advanced work of Peixoto and Oort [4] detail the basic physics: the actual problems encountered in constructing climate models are summarized in Trenberth [5]. We concentrate solely on certain aspects of the energy transfer at a basic level; more sophisticated treatments do exist [6–9]. Moreover, this paper deals exclusively with a representation of observational material that can be useful because it imposes constraints on assumptions used in the treatment of transverse transport of latent energy in more detailed climate models.

Despite the complexities of the real atmosphere, the global annual average temperature of 288 K (ref. [3], p. 2) may be estimated with remarkably few modifications of the classic homogeneous black-body model of Earth [3,10–13]. The present paper extends our earlier model [11], in which the ground and atmosphere, for the entire Earth, were represented by two radiating heat reservoirs that could exchange energy. Both could radiate into space and both received energy from the Sun. Energy balance was achieved by introducing one non-radiative flux $S_{NR}$ (see Sec. II, below).

In this paper, instead of treating the entire earth as a single entity, we replace its surface (and atmosphere) by a covering set of *non-interacting* cells. In each cell, vertical energy balance, with the addition of one non-radiative source, is specified by the procedure of [11]. The 864-cell division is that used in a detailed professional climate model [14]. For each cell we used observed mean values of surface reflectivity, mean cloudiness, and insolation appropriate to the latitude and time of year.

There are two approaches. With a simple assumption for the form of $S_{NR}$, one finds a reasonable representation of the observed surface temperature [15]. This is a simplification of the extensive parameterization used by, *e. g.*, Budyko [7] for representing cloud cover, evaporation rates, release of latent energy, etc. Alternatively if one uses the observed surface temperatures one finds a close dependence of $S_{NR}$ on latitude and time of year. The term $S_{NR}$ includes general transverse transfer of energy from equatorial cells to polar region cells.

## II. THE BASIC MODEL EQUATIONS

We first briefly review the two-layer model which makes use of the fact that the incoming solar radiation (at ~500 nm) and the outgoing terrestrial radiation (at ~ 10 μm) occupy distinct spectral regions. We refer to these as "solar" and "IR," respectively (for details of the spectral distributions see [11], Fig. 2, and [4], Figs. 6.1, 6.2). The atmosphere layer (a heat reservoir), at temperature $T_A$, has a solar reflectivity and absorptivity $r_A$ and $a$, respectively, and an IR absorptivity $\varepsilon$. The symbols $a$ and $\varepsilon$ are used here in place of $f$ and $g$ in the original paper but all other notation is preserved. The surface layer (also a heat reservoir), at temperature $T_E$, has a solar reflectivity $r_S$ and is assumed to absorb all incident IR. The solar radiation has a flux $S_0 = 342$ Wm$^{-2}$ when averaged over time and over the surface of the earth. When the solar radiation hits the upper layer, a fraction $r_A$ is reflected and the fraction that enters the layer is $(1-r_A)$. Of this, $a(1-r_A)$ is absorbed. In this first pass the surface therefore receives $(1-r_A)(1-a)$, of which $r_S$ is reflected, leaving $(1-r_A)(1-a)(1-r_S)$ as the total fraction of original incident solar flux to be absorbed at the surface. Following this logic, a diagram can be constructed showing all the fractions of absorbed and reflected radiation including the term $S_{NR}$ for non-radiative upward energy transfer (Fig. 1) for each cell. When multiple reflections are included, a factor $k_M = (1 - r_A\, r_S)^{-1}$ appears in terms involving reflections between the surface and atmosphere. The non-radiative flux $S_{NR}$, which served as an arbitrary model parameter in the global calculation, will be seen to play a much greater role in the current work; it shows that horizontal transport of energy is effective.

The energy balance equations for the upper and surface layer take the form

$$2\varepsilon S_A - \varepsilon S_E = AS_0 + S_{NR} \tag{1a}$$

and

$$-\varepsilon S_A + S_E = BS_0 - S_{NR}, \tag{1b}$$

respectively. These equations express net incoming solar and non-radiative flux on the right hand sides and net outgoing IR flux on the left. Here $S_A$ and $S_E$ are defined as the ideal Stefan-Boltzmann fluxes $S_A = \sigma T_A^4$ and $S_E = \sigma T_E^4$, where $\sigma = 5.67$ x $10^{-8}$ Wm$^{-2}$K$^{-4}$. The quantities $A$ and $B$ correspond to the fractions of $S_0$ ultimately absorbed by the atmosphere and surface, respectively,

$$A = a(1-r_A) + ak_M(1-r_A)^2(1-a)r_S \tag{2a}$$

and

$$B = k_M (1 - r_A)(1 - r_S)(1 - a).$$ (2b)

Global and time averages of all the parameters are inserted and equations (1) are easily solved for $S_A$ and $S_E$, from which temperatures $T_A$ and $T_E$ are then obtained. This original model was used to investigate broadly the effect of non-solar-related energy sources at the surface.

## III.  DETAILS OF THE CELLULAR MODEL

## A.  Rationale and design

The model is now extended by averaging surface features and insolation for individual cells.  We choose the grid scheme used by Hansen *et al.* [14] in which the earth's surface is divided into 864 cells of dimension 8º x 10º (latitude by longitude). For each of the 864 cells, the land fraction and the annually averaged parameters for observed total cloud cover fraction, calculated incoming annual average solar radiation, and land and sea reflectivities are stored. We then assume that Equations (1) are satisfied within each cell independently of other cells [16].   The land fraction $f_{land}(n)$ is taken directly from Hansen *et al.* [14].  It is shown by contours in Fig. 2, which establishes the scale and resolution of subsequent maps.  The 18-year average annual cloud fraction coverage was taken from data collected by satellites from 1983 to 2001 under the International Satellite Cloud Climatology Project [17].  Similarly, surface reflectivities ($r_S$) were estimated by using the ISCCP surface reflectivities in coordination with Table 4.2 in Hartmann [3].  Hartmann specifies albedo ranges and typical values for distinct land types. Table I shows the land types, albedo ranges, and typical albedo values. The estimates of $r_S$ are shown in Table II. All parameters are determined at the center of the cell and the values applied to the entire area of the cell. The energy balance equations are then solved at each location. When results are compared to global values, the local values are weighted by fractional cell area and summed. Our model does not take into account changes in atmospheric components, the parameters' temperature dependence [13], or any geothermal variations.

## B. Basic astronomy

The value of a parameter $g$ that depends on the choice of cell will be written $g(n)$, where unless otherwise specified, $n$ is an arbitrarily assigned cell number. To calculate the incoming solar radiative flux to a cell, we use astronomical definitions and notations to track the sun's position relative to a cell's midpoint throughout the year. The incident radiation is the solar radiation at the distance to the earth, $1368 \text{ Wm}^{-2}$ multiplied by the sine of $H(n)$, where $H(n)$ is the angle of the sun from the horizon of the surface area in cell $n$,

$$S_0(n) = 1368 \sin H(n,t,t') \quad (\text{Wm}^{-2}). \tag{3}$$

Here (see Appendix and [3], pages 29-31 and 347-349)

$$\sin H(n,t,t') = \sin\delta(t')\sin\beta(n) + \cos\delta(t')\cos\beta(n)\cos t, \tag{4}$$

where $\beta$ is the latitude, $\delta$ is the sun's declination angle, $t'$ is the time elapsed since the vernal equinox, and $t$ is the hour angle. During the average over 24-hour periods, the angle $t$ is limited by sunrise and sunset conditions, *i. e.*, there is no contribution when $\sin H$ is negative. The annual average of $S_0$ as a function of latitude computed from Eq. (3) is plotted in Fig. 3, along with the observed $S_E$ and $T_E$. Eq. (3) is adequate for the purposes of our average annual model. For more precise work, monthly averages would require a slight correction of at most ±3% resulting from Earth's orbital eccentricity. The annual average solar irradiance ranges from a minimum of 176 Wm$^{-2}$ at the poles to 416 Wm$^{-2}$ at the equator, with a global annual average of 342 Wm$^{-2}$.

## C. Other cell parameters

Since the solar reflectivities and the absorptivities of the cloud layer vary from cell to cell, the solar parameters $A$ and $B$, Eqs. (2a,b), also become cell-dependent. There are three relevant cell-varying parameters, the solar reflectivities and the absorptivity of the cloud layer. The surface reflectivity $r_S(n)$ of cell $n$ is taken from sea and land reflectances (see Table I) weighted by the corresponding surface type fractions,

$$r_S(n) = f_{\text{land}}(n)r_{\text{land}} + [1 - f_{\text{land}}(n)]r_{\text{sea}}. \tag{5}$$

This formula was used only in high latitude regions (|lat| ≥60 degrees). Remaining reflectivities came from ISCCP data [17,18]. The atmospheric reflectivities are similarly found to be

$$r_A(n) = f_{\text{cloud}}(n)r_{\text{cloud}} + [1 - f_{\text{cloud}}(n)]r_{\text{clear}}, \tag{6}$$

where the clear air albedo ($r_{clear}$) can be assumed as 0.15 (Ref. [3], page 75). The cell value of the planetary albedo $\alpha_P(n)$ is the fraction of original unabsorbed incident solar radiation that leaves the system. On our model (see [11], eq. B3) it is given by

$$\alpha_P(n) = r_A(n) + k_M(n)[1 - a(n)]^2 [1 - r_A(n)]^2 r_S(n), \qquad (7)$$

which depends upon two unknowns, $a(n)$, the atmospheric solar absorptivity, and $r_{cloud}$ (through $r_A$, eq. 6). For absorptivity, the expression corresponding to (6) is

$$a(n) = f_{cloud}(n)a_{cloud} + [1 - f_{cloud}(n)]a_{clear}. \qquad (8)$$

If we assume that clear air absorbs no solar radiation, we have

$$a(n) \cong f_{cloud}(n)a_{cloud}. \qquad (9)$$

There then are two whole-planet parameters left to specify, $r_{cloud}$ and $a_{cloud}$. To estimate then, we impose a set of reasonable planetary albedo values (ref. [3], Fig. 2.9a, page 33) as a constraint upon Eq. (7). Values of $r_{cloud}$ and $a_{cloud}$ were varied at intervals of 0.01 until we had the greatest number matches with the known $\alpha_P(n)$. With this bare minimum of free parameters, a match to observed values occurred in 711 of the 864 cells. Most of the remaining cells, having more than 15% difference from the Hartmann values, are near the poles. The result is $r_{cloud}$ = 0.27 and $a_{cloud}$ = 0.06, in reasonable agreement with the parameters found in [11]. These produce not only a good localized match but also give the commonly accepted global planetary albedo of 0.30 when averaged. Once adopted, these values of $r_{cloud}$ and $a_{cloud}$ are not changed in the course of the calculations.

The cellular IR absorptivity $\varepsilon(n)$, which is also the cellular IR emissivity, is taken to have the same form as the solar absorptivity, Eq. (8). Parameters used were $\varepsilon_{cloud} = 1$ and $\varepsilon_{clear} = 0.90$.

## IV. APPLICATION OF THE MODEL

It is important to realize that $S_{NR}$ really consists of three terms: a true vertical transport of non-radiative energy, a contribution from heat flowing in from neighboring cells, and accumulation of systematic errors. One should not conclude that we have neglected lateral transport of energy into each cell because it does not appear as another term in the equations. Actually a portion of the lateral heat flows will be effectively redirected into a vertical heat flux; the quantity $S_{NR}$ will include this contribution. Consider, for example, a horizontal wind, carrying water vapor from one cell to the next. When the vapor precipitates in the form of rain, it releases latent energy. That energy release contributes to the vertical energy balance in the (receptor) cell. This energy enters "horizontally" into the cell. Similarly, in some nearby cell, some vertical energy flux went into evaporation, and that cell suffers an energy loss if the water vapor is transported out of the cell. Since our cells are large, we do not expect the horizontal energy transport to cover more than one or two cells, for local disturbances. Consequently there may be a correlation between a loss of $S_{NR}$ in one cell and a gain in $S_{NR}$ in a nearby cell. The annual latitude dependence of $S_{NR}$, which will be discussed in Sec. V, may well be made up of such transport contributions. Finally, the radiative transfer in the model used is greatly simplified and it would not be surprising for systematic errors to be introduced because of it; all systematic errors end up in the $S_{NR}$ term. Another systematic error may arise from our assumption that the theory continues to work at latitudes having long periods of low solar irradiance, when cloud cover may vary significantly between polar winter and summer periods.

## A. Predicting $T_E(n)$

We now generalize Eqs. (1a,b) to the cellular case:

$$2\varepsilon S_A(n) - \varepsilon S_E(n) = A(n)S_0(n) + S_{NR}(n) \tag{10a}$$

and

$$-\varepsilon S_A(n) + S_E(n) = B(n)S_0(n) - S_{NR}(n). \tag{10b}$$

$A(n)$ and $B(n)$ are given by equations identical to (2a,b) in which certain parameters are made cell dependent, as discussed above.

There is a class of models in which $S_{NR}$ is assumed, given as a function of other atmospheric parameters. In this case, we refer to these applications of the model as "$T_E$-predictive." For example, we allowed the annual average value of $S_{NR}(n)$ to depend on land and sea fraction and took it to be proportional to the solar input, as follows:

$$S_{NR}(n) = [0.03 f_{\text{land}}(n) + 0.16 f_{\text{sea}}(n)] \cdot S_0(n). \tag{11}$$

From the solution of each pair of Eqs. (10a,b) for $S_E(n)$, $T_E(n)$ was calculated on the basis of the assumed $S_{NR}(n)$. The numerical coefficients in Eq. (11) were chosen by an extensive search of parameter space to produce the observed global average temperatures ($T_E = 288$ K, $T_A = 250$ K). This flux has a global average of 42.3 $Wm^{-2}$, an improvement on the one-dimensional model [11] in that the latter had been unable to accommodate *any* non-zero average $S_{NR}$ without compromising other assumed input parameters. We emphasize that the only completely arbitrary parameters in the fit were the two numerical coefficients in Eq. (11). We consider this remarkable; it endorses the general reasonableness of the elementary two-temperature model for individual cells. The form of Eq. (11) also suggests a global asymmetry in the distribution of $S_{NR}$. We return to this later.

The preliminary cell results are shown in Fig. 4a and are compared with measured values (satellite data [17]), Fig. 4b. These diagrams show how a set of calculated locally-determined temperatures (Fig. 4a) having the correct global average may disagree significantly from observed local values (Fig. 4b) having the same global average. While this is not the least bit surprising, a comparison of the two parts of Fig. 4 provides a qualitative evaluation of the errors that occur in the making of simple models. Comparison of the two panels shows the calculated temperatures are ~5–10 K too high in equatorial regions and ~5–10 K too low in the temperate zones. Considering the simplicity of the model, the agreement is quite good, but the values of the two needed parameters are far from being uniquely determined.

## B. Predicting $S_{NR}(n)$

The linear equations (1a,b or 10a,b) lend themselves equally to computing any two of the quantities $S_{NR}$, $S_A$, and $S_E$, given the third one and $S_0$ as an input.

Therefore, we may take $T_E(n)$ as known input parameters from ISCCP satellite data [17] and $S_0(n)$ from Eq. (3), and use the balance equations to calculate $S_A(n)$ and $S_{NR}(n)$. This "$S_{NR}$-predictive" mode of calculation is accomplished most easily by combining the two flux equations (10a,b),

$$S_{NR}(n) = [A(n) + 2B(n)]S_0(n) - (2 - \varepsilon)S_E(n).\qquad(12)$$

Recall that $S_E(n) = \sigma T_E(n)^4$. Eliminating $S_E$ from (10a) and (10b) results in the companion equation for $S_A$:

$$\varepsilon(2 - \varepsilon)S_A(n) = [A(n) + \varepsilon B(n)]S_0(n) + (1 - \varepsilon)S_{NR}(n).\qquad(13)$$

The results for the annual average of $S_{NR}(n)$, calculated from the observed surface temperature $T_E(n)$, are shown in Figure 5a. $S_{NR}(n)$ and $S_A(n)$ have global averages of 64 and 236 $Wm^{-2}$ (the latter corresponding to $T_A = 254$ K), respectively. In the earlier non-cellular model [11], the highest value of $S_{NR}$ that could be obtained without unreasonable parameters was 40 $Wm^{-2}$, and in the above $T_E$-predictive mode it was 42.3 $Wm^{-2}$, so the $S_{NR}$-predictive mode result 64 $Wm^{-2}$ represents a further improvement. The generally quoted global average of $S_{NR}$ is 102–105 $Wm^{-2}$ [19,20]. $S_{NR}(n)$ appears to be most negative at the higher latitudes and most positive near the equator. A negative value of $S_{NR}(n)$ corresponds to non-radiative energy transfer from the atmospheric layer to the surface layer or a lateral flow into the cell, as discussed earlier.

A distinctive feature of our results is the prominent drop of $S_{NR}$ in the regions of the Sahara and Saudi Arabia, Figs. 5a and 5b, at matrix elements (20–25, 14–16). Its cause is a confluence of strong effects on the two terms in Eq. (12): relatively high surface reflectivity and low cloud cover, which reduce the first term, and relatively high temperature, which increases the absolute value of the (negative) second term. A similar but milder dip appears in the eastern region of Australia. Aside from these anomalies, the residuals are generally small (see below).

## C.  The zonal average of $S_{NR}$

"Zonal" averages are made over cells lying within zones having the same latitude. Following convention [21], we denote zonal averages by angular brackets $\langle ... \rangle$. If the area of cell $n = (p,q)$ is $A(p,q)$, where $p$ is the latitude cell index and $q$ is the longitude cell index, we have , for example,

$$\langle S_{NR}(p) \rangle = \frac{1}{A(p)} \sum_q A(p,q) S_{NR}(p,q), \qquad (14)$$

where $A(p)$ is the total area of zone $p$,

$$A(p) = \sum_q A(p,q). \qquad (15)$$

For convenience the latitude index $p$ will be converted into the latitude $\beta$, measured in degrees, at the center of the cell and we will write, again for example, $S_{NR}(\beta) = \langle S_{NR}(p) \rangle$. The zonal averages $S_{NR}(\beta)$ are shown explicitly in Figure 6, where a very regular latitude dependence emerges clearly. Indeed, $S_{NR}(\beta)$ can be represented to within ±9 Wm$^{-2}$ by

$$S_{NR}^{\text{fit}}(\beta) = 40 + 80 \cos 2(\beta + \Delta) - 10 \sin 6 \, | \beta |, \qquad (16)$$

with $\Delta = 5$ degrees. In Figure 6 the residuals between $S_{NR}(\beta)$ and its fit are also shown. When the zonal average is removed from the cellular results shown in Fig. 5a, the residuals shown in Fig. 5b are obtained. The zonally averaged residuals are also shown in Fig. 6.

Both the general smoothness and the relatively small size of the residuals, on the order of 5–10% of the solar input, establish the fact that Eq. (16) is not simply an average but a useful average. It probably represents both latitude gradient in the amount of vertical latent heat transport (more evaporation and condensation in equatorial regions) and horizontal transport of energy from equatorial to polar regions.

It is not surprising that $S_{NR}(\beta)$ has an asymmetry between the northern and southern hemispheres. In retrospect we see that the *ad hoc* form, Eq. (11), used in the $T_E$-predictive calculation, was biased toward the southern hemisphere where the sea fraction is dominant. Figure 7 shows the 18-year average surface reflectivity in months 1 and 7 [17], giving further insight into the peculiarities of $S_{NR}$. The reflectivities are slightly higher in the northern temperate zone than in the southern. In the polar regions, the times of greater solar irradiance may not occur when the surface reflectance is at its average value. Indeed, the polar regions are generally rather anomalous. We have not concerned ourselves too much with them because the model, generally limited to dealing with annual averages, lacks the ability to describe accurately the effect of Arctic and Antarctic nights in which $S_0(n) = 0$. There is some seasonal variation in the cloud cover data and this may be responsible for the fact that the interesting term $-10\sin6|\beta|$ does not hold in the polar regions. One is tempted to speculate on the origin of the sinusoidal term: the insolation does not have a pure $\cos\beta$ dependence because of the inclination of Earth's axis to the plane of its orbit; and Hadley cells [22] may play a role in it.

The existence of extensive databases provides the student an opportunity to explore many other effects through the medium of this theory. As an example, we have used data from ISCCP [17] consisting of averages of the measured parameters over an 18-year period at each month. For each month, the average $S_{NR}(\beta,t)$ was calculated and compared to $S_{NR}(\beta)$ by looking at the difference between the two. This difference also appears to follow a trend that is most clear in the region between latitudes –60° and +60°. For this region the difference is approximately linear and oscillates about $\beta = 0°$ with a period of one year. The difference itself can be fitted well to

$$\Delta S_{NR}(\beta,t) = -285 \cdot (\beta/60°) \cdot \sin(30° t) \tag{17}$$

(where: $t = 0$, September 15; $t = 1$, October 15; etc.)

Now, we have shown that the annual average $S_{NR}$ follows Eq. (16), and the difference between monthly and annual values follow Eq. (17), so the monthly $S_{NR}$ can be written as

$$S_{NR}^{\text{fit}}(\beta,t) = 40 + 80\cos 2(\beta + \Delta) - 10\sin 6 |\beta| + \Delta S_{NR}(\beta,t) \tag{18}$$

in the specified region $-60° \leq \beta \leq +60°$. For the month of January, the values of $S_{NR}$ calculated directly from the data are compared to those given by Eq. (18) in Figure 8. The fitting is very close to the calculated values in the region $-30° \leq \beta \leq +45°$. The discrepancies of up to 50 Wm$^{-2}$ outside this region are most likely due to the hemispheric asymmetry of $S_{NR}$. This asymmetry was ignored in Eq. (17) which is antisymmetric about $\beta = 0°$ and uses a perfectly sinusoidal maximum value for $\beta = \pm 60°$. The calculations from data, however, show that during the southern summer, $S_{NR}(-60°,t)$ increases to close to 400 Wm$^{-2}$, but in the northern summer $S_{NR}(+60°,t)$ does not even reach 300 Wm$^{-2}$. Also, at any month, the value of $S_{NR}(60°,t)$ does not equal $S_{NR}(-60°,t)$, as in Eq. (17), leading to an over- or under-estimation of $S_{NR}$ from Eq. (18).

Finally, in Figure 9 we compare averages of $S_{NR}$ over two months in succession (February and March 1995). The 18-year average for the respective months has been subtracted and the polar regions have been omitted from the diagram because average monthly variations in reflectivities and cloud cover may be large. Deviations of the order of 50 Wm$^{-2}$ show the presence of large-scale persistent non-radiative "weather systems," each of which is composed of about 20 independent cells. These graphs show how the expected horizontal transport of energy in the east-west direction is effected. These systems produce deviations in $S_{NR}$ much larger than those shown in Fig 5b and therefore are real. Only when averages are taken over the 18-year database do the smooth results given by Eqs. (17) and (18) and Figs. 6 and 8 result, with the evidence of the east-west transport being averaged out. These graphs are important in that they show horizontal east-west transport of energy can be included in our model, providing small enough time intervals are used.

## V. SUMMARY AND DISCUSSION

The two-level global model of paper I has been applied locally, that is, each cell in a grid has been assumed to have an annual average temperature and the fluxes have been determined by the cell's own parameters and average insolation. The observed local planetary albedo is used as a control on the modeled surface and atmospheric reflectivities. In the model's more successful ($S_{NR}$-predictive) implementation the set of surface temperatures is used as input; the nonradiative flux from the surface $S_{NR}$ and the ideal atmospheric radiative flux $S_A$ are the principal outputs. The globally averaged $S_{NR}$ is predicted to be about 64% of its usually quoted global value of 102–105 $Wm^{-2}$, an improvement over $T_E$-predictive models using assumed non-radiative fluxes, where values of only 0–40% were possible.

The value of the globally averaged atmospheric radiative flux $S_A$ is 236 $Wm^{-2}$, adequate to maintain overall radiative balance with an effective atmospheric radiative temperature of $T_A = 254K$) (recall that the emissivity is taken as 0.89, as in paper I). A drawback of our one-temperature atmosphere is that it is constrained to predict that the downward IR flux is identical to the upward flux. In the real atmosphere, a temperature gradient exists and the lower layers most effective in radiating downward are at a higher temperature. The downward flux should thus be greater than $\varepsilon S_A$, as is observed [19]. We are developing a three-temperature version of our model, to be the subject of paper III in the series, using a radiative transfer model for the atmospheric structure; according to preliminary estimates an appropriately larger value of $S_{NR}$ will be obtained that will balance the extra downward IR.

Having noted a correlation between $S_{NR}(n)$ and latitude, we examined its zonal average. There resulted a clear asymmetry between the northern and southern hemispheres, illustrating the effect of asymmetry of the hemispheric land masses and the differences in reflectivity parameters resulting from the nature of the Arctic and Antarctica. Fittings of other data sets to develop formulae for individual components of $S_{NR}$ are discussed by Budyko [7].

We emphasize the accessibility of both the data and our model representations to students interested in applying the elementary aspects of climatology to real data. As suggestions for future workers: (1) It would be interesting to use data bases averaged over different periods of time to see if long-term differences in the form of Eqs. (17) and (18) result. (2) A study of the form, structure and persistence of these large scale monthly "weather patterns" during an El Niño cycle should be informative.

## ACKNOWLEDGMENTS

We are grateful to Profs. D. Hartmann and K. Trenberth for valuable correspondence and to Karen Kiselycznyk of the Laboratory for Laser Energetics for her expert assistance with the illustrations. This work was supported in part by NSF (REU) grant PHY 99-87413.

# APPENDIX A.  SPHERICAL ASTRONOMY FUNDAMENTALS

The altitude $H$ of the sun can be found by applying the law of cosines [23] to the observer's spherical triangle $\Delta ZNS$ where $Z$ is the observer's zenith, $N$ is the north celestial pole, and $S$ is the sun.  Then arc $ZS$ is $90° − H$;  arc $NZ$ is $90° − \beta$, where $\beta$ is the latitude, and arc $NS$ is $90° − \delta$, where $\delta$ is the *declination* of the sun, available from a table look-up in (*e.g.*)[24].

The angle $\angle SNZ$ is called the Sun's *hour angle $H$*; instead of being measured in degrees it is measured in time units from 12:00 noon (1 hour = 15°). It is negative (positive) when the Sun is in the eastern (western) half of the sky. These quantities are related by the spherical law of cosines,

$$\sin H = \sin\beta \sin\delta + \cos\beta \cos\delta \cos t , \qquad (A1)$$

At sunrise, $H = 0$; the equation determines the time at sunrise $t = −t_0$.  At sunset, again $H = 0$ and $t = +t_0$.  The length of day is then $2t_0$.

The declination of the sun can also be approximated.  The sun moves along a great circle, called the *ecliptic*, which is inclined at an angle $i = 23.44°$ to the celestial equator.  At the vernal equinox (~March 21) the sun is at a point $V$, one of the two intersections of the celestial equator and the ecliptic, and is moving from negative to positive declination.  Let the point $P$ be on the celestial equator, with arc $NSP = 90°$.  Consider the spherical triangle $VSP$.  The angle $\angle SVP = \iota$ and the angle $\angle SPV = 90°$.  The arc $VS$ is *approximately* $\Omega t'$ where $\Omega = 360°/1\text{yr}$ and $t'$ is the time elapsed since the vernal equinox.  From the spherical law of sines for $VSP$ one has:

$$\sin\delta = \sin i \; \sin\Omega t' , \qquad (A2)$$

Since the Sun moves slightly faster (slower) on the ecliptic than average when we are at perihelion, January (aphelion, July), this is only an approximate relation.

## List of symbols and abbreviations

Symbols of the form $T_A(n)$ are not included. The meaning of such a symbol is "the value of $T_A$ in cell $n$."

| | |
|---|---|
| $A$ | Solar atmospheric input parameter, Eq. 2a |
| $A(p)$ | Area of a band of cells of latitude index $p$ |
| $A(p,q)$ | Area of a cell of latitude index $p$ and longitude index $q$ |
| $a$ | Atmosphere's absorptivity of solar radiation |
| $a_{clear}$ | Absorptivity of solar radiation in cloudless air |
| $a_{cloud}$ | Absorptivity of solar radiation in air with clouds |
| $B$ | Solar surface input parameter, Eq. 2b |
| $f_{cloud}$ | Fraction (of a cell area) consisting of cloud |
| $f_{land}$ | Fraction (of a cell area) consisting of land |
| $H$ | Horizon angle of the sun |
| IR | Refers to that part of the spectrum with wavelengths longer than 600 nm; largely not absorbing the sun's spectrum but absorbing much of Earth's |
| $k_M$ | Multiple reflection parameter |
| $n$ | Cell label, also in matrix style $p, q$ |
| $p, q$ | Latitude and longitude cell indexes, respectively |
| $r_A$ | Reflectivity of solar radiation by the atmosphere |

*(List of symbols, continued)*

| | |
|---|---|
| $r_{\text{clear}}$ | Reflectivity of solar radiation by clear air |
| $r_{\text{cloud}}$ | Reflectivity of solar radiation by the cloud portion of a cell |
| $r_{\text{land}}$ | Reflectivity of solar radiation by the land portion of a cell |
| $r_S$ | Reflectivity of solar radiation by the surface |
| $r_{\text{sea}}$ | Reflectivity of solar radiation by the sea portion of a cell |
| $S_0$ | Solar constant averaged globally and over time, 342 Wm$^{-2}$ |
| $S_A$ | Ideal radiative flux in the atmosphere or upper atmosphere layer, $\sigma T_A^4$ |
| $S_E$ | Ideal radiative flux in the surface layer, $\sigma T_E^4$ |
| $S_{NR}$ | Net non-radiative flux upward from the surface |
| $t$ | Solar hour angle (see Appendix) |
| $t'$ | Time elapsed since the vernal equinox |
| $T_A$ | Effective radiative temperature of the upper model layer, representing that of the atmosphere |
| $T_E$ | Temperature of the lower model layer, representing that of the surface of Earth |
| $\alpha_P$ | Planetary albedo |
| $\beta$ | Latitude associated with a set of cells |
| $\delta$ | Solar declination angle |
| $\Delta$ | Fitting parameter (phase shift); see Sec. IV |

*(continued)*

*(List of symbols, continued)*

$\varepsilon$            Atmosphere's absorptivity (and emissivity) in the IR

$\varepsilon_{\text{clear}}$          Clear air absorptivity (and emissivity) in the IR

$\varepsilon_{\text{cloud}}$         Cloud absorptivity (and emissivity) in the IR

$\sigma$            Stefan-Boltzmann constant, $5.67 \times 10^{-8} \, \text{Wm}^{-2} \, \text{K}^{-4}$

Table I. Land types and their associated albedo ranges (in percentages) deduced from a map, Fig. 5.14, by Dickinson [21] and Table 4.2 of Hartmann [3].

| Land type (Dickinson) | Land type (Hartmann) | Albedo range | Typical value |
|---|---|---|---|
| Tundra & desert | Dry soil/desert | 20-35 | 30 |
| Grass & shrub | Short green vegetation | 10-20 | 17 |
| Crop | Dry vegetation | 20-30 | 25 |
| Wetland & irrigated | Short green vegetation | 10-20 | 17 |
| Evergreen tree | Coniferous forest | 10-15 | 12 |
| Deciduous tree | Deciduous forest | 15-25 | 17 |

Table II. Corrected values for land and sea reflectivities in the high latitudes (see text). A negative latitude corresponds to the southern hemisphere. For –76° to –89° $r_{land}$ is set at 0.6 because of the year-round Antarctic ice. The latitude –68° reflectivity is set slightly lower than Antarctica as a result of a lack of permanent ice. In latitudes 68° to the north pole , $r$ gradually increases, taking seasonal snow and ice into account. $r_{sea}$ at 60° and –60° is set at 0.3 to avoid a sharp jump from water set at 0.1 in the mid-latitudes to the higher polar values for ice and snow cover. These values are rough estimates for partial and seasonal snow and ice cover. Fresh snow can have an albedo up to 0.9, old, melting snow up to 0.65, and sea ice without snow cover up to 0.4 ( see [3], table 4.2, p. 88). Values not shown (- - -) are longitude-dependent and are taken directly from satellite data in detail.

| Latitude (deg) | –89 | –84 | –76 | –68 | –60 | –52 to 52 | 60 | 68 | 76 | 84 | 89 |
|---|---|---|---|---|---|---|---|---|---|---|---|
| $r$ (land) | 0.6 | 0.6 | 0.6 | 0.55 | - - - | - - - | - - - | 0.4 | 0.5 | 0.55 | 0.6 |
| $r$ (sea) | 0.6 | 0.6 | 0.6 | 0.55 | 0.3 | - - - | 0.3 | 0.4 | 0.5 | 0.55 | 0.6 |

## References

1.  J. Hansen, M. Sato, R. Ruedy, A. Lacis, and V. Oinas, Global warming in the twenty-first century:  an alternative scenario, Proc. Natl. Acad. Sci. (US) 97, 9875-9880 (2000)

2.  R. M. Goody and J. C. G. Walker, Atmospheres (Prentice-Hall, Inc., Englewood Cliffs, NJ, 1972)

3.  D. L. Hartmann, *Global Physical Climatology* (Academic Press, San Diego, 1994)

4.  J. P. Peixoto and A. H. Oort, *Physics of climate* (American Institute of Physics, New York, 1992)

5.  K. E. Trenberth, ed., *Climate System Modeling* (Cambridge University Press, Cambridge, UK, 1992)

6.  M. I. Budyko, The effect of solar radiation variations on the climate of the Earth, Tellus 21, 611-619 (1969)

7.  M. I. Budyko, *The Earth's Climate: Past and Future* (Academic Press, New York, 1982).  English translation by the author of the original Russian edition of 1980.

8.  K. E. Trenberth, J. M. Caron, and D. P. Stepaniak, The atmospheric energy budget and implications for surface fluxes and ocean heat transports, Climate Dynamics 7, 259-276 (2001)

9.  K. E. Trenberth and J. M. Caron, Estimates of meridional atmosphere and ocean heat transports, J. Climate 14, 3433-3443 (2001)

10. C. Kittel and H. Kroemer, *Thermal Physics* (W. H. Freeman and Co., New York, ed. 2, 1980), fourth printing, pp. 115-116

11. R. S. Knox, Physical aspects of the greenhouse effect and global warming. Amer. J. Phys. 67, 1227-1238 (1999), to be referred to as paper I.

12. S. Arrhenius, On the influence of carbonic acid in the air upon the temperature of the ground. Phil. Mag. [ser. 5] 41, 237-276 (1896)

13. J. R. Barker and M. H. Ross, An introduction to global warming. Amer. J. Phys. 67, 1216-1226 (1999)

14. J. Hansen, G. Russell, D. Rind, P. Stone, A. Lacis, S. Lebedeff, R. Ruedy, and L. Travis, Efficient three-dimensional global models for climate studies: models I and II. Monthly Weather Rev. 3, 609-662 (1983)

15. To avoid possible confusion as to what the slightly ambiguous term "surface temperature" means, for us it always means that temperature given by the data set used.

16. All programming is done using MATLAB™, an array-based language with simple commands. In the MATLAB environment, each computation is performed simultaneously on each cell in an identical manner. A very useful source for MATLAB programming styles is the Appendix of G. J. Borse, *Numerical methods with MATLAB®: a resource for scientists and engineers* (PWS Publ. Co., Boston, 1997)

17. C. Brest (technical contact), ISCCP-D2 Monthly Means and Climatology data sets. Retrieved 21 Jan 2002 from International Satellite Cloud Climatology Project website: http://isccp.giss.nasa.gov/products/browsed2.html (verified 18 Sep 2002)

18. The ISCCP data for $r_S$ was not used in regions above ± 52° latitude, because the data sometimes are greater than unity as a result of glaring and the angle of the satellite to the northern and southern polar regions.  This being inconsistent with our definition of a reflection coefficient, the surface reflectivities in these regions were estimated from the Hartmann [3] table 4.2.

19. J. T. Kiehl and K. E. Trenberth, Earth's annual global mean energy budget, Bull. Amer. Meteorol. Soc. 78, 197-208 (1997)

20. See Budyko [7], page 70.

21. R. E. Dickinson, Land surface, in Trenberth, ref. [5], Chap. 5, pp. 149-171

22. See Hartmann [3], pp. 152-154.

23. W. M. Smart, *Text-Book on Spherical Astronomy* (Cambridge University Press, Cambridge, 1947);  L. G. Taff, *Computational Spherical Astronomy* (John Wiley and Sons, New York, 1981).

24. R. Gupta, ed., *Observer's Handbook 2002* (The Royal Astronomical Society of Canada, Toronto, 2002)

# Figure captions

1. Movement of radiative and non-radiative energy in the two layer schematic. Horizontal arrows indicate deposition in the layer. Lighter arrows represent radiation that is either reflected, or passed through a layer unabsorbed. Multiple reflections of solar radiation are accounted for by the factor $k_M$ in Eqs. (1a) and (1b) in the text. The factors $\alpha_p$, $A$, and $B$ are the overall fractions of $S_0$ that are reflected, absorbed by the upper payer, and absorbed by the lower layer, respectively. Other symbols as defined in the text.

2. Global land fraction plotted with contours at 0.25, 0.5, and 0.75. This map may be used as a template in the study of Figs. 4, 5, and 7. It also gives a good indication of the resolution afforded by our 864-cell calculations. Horizontal axis: ticks correspond to centers of 10° cells located (centered) at longitudes −180°(1), −170°(2), ..., 0°(19), ..., +170°(36). Vertical axis: ticks correspond to centers of 8°cells at latitudes −84°(2), −76°(3), ..., −4°(12), +4°(13), ..., +84°(23). The bottom and top rows correspond to centers of 2° cells at −89°(1) and +89°(24), respectively.

3. Average annual insolation (light solid curve and right-hand scale), outward infrared flux from the surface (dashed curve, right-hand scale), and surface temperature (heavy curve, left-hand scale) as a function of latitude. The latter two are latitude averages based on satellite temperature data [17].

4. (a) Computed "$T_E$-predictive mode" surface temperatures (in K) using a direct cellular extension of the elementary model of reference [11], as described in the text. The average non-radiative flux is 40 Wm$^{-2}$ and the average surface temperature is 288 K. (b) Observed surface temperatures, defined as those obtained by satellite [17], with an average of 288 K. See the caption of Fig. 2 for the key to the axes. In both of these plots, most of the regional variation in surface temperature comes from the latitudinal variation of insolation and from the geographic variation of albedo.

5. (a) Computed "$S_{NR}$-predictive mode" non-radiative flux $S_{NR}(n)$, in Wm$^{-2}$, using the cellular model but with surface temperatures as input. (b) $\delta S_{NR}(n)$, the residue after subtracting the annual average value of $S_{NR}(n)$ as a function of latitude (see Eq. 17). The region of large residuals corresponds to the Sahara desert region. See the caption of Fig. 2 for the key to the axes.

6. Distribution of $S_{NR}$ by latitude. Squares represent the "experimental" values based on our model, and diamonds are the numerical fit, Eq. (16). Triangles are the residuals.

7. Comparison of surface reflectivities in the months of January (a) and July (b). The asymmetry between the northern and southern hemispheres seen in many climatological studies can be appreciated from the variability in the north polar region and the near-invariance in the south polar region. See the caption of Fig. 2 for the key to the axes.

8. A sample determination of nonradiative flux $S_{NR}$ for the month of January (18-year average). Diamonds: values determined from our model calculation. Squares: value determined from the fitting function, Eq. (18). The fit is meant to be valid for latitudes satisfying $-60° \leq \beta \leq +60°$.

9. (a) The monthly values of $S_{NR}$ for February 1995 minus the 18-year average value of $S_{NR}$ for February, for the temperate zones. Large deviations, $\sim$50 Wm$^{-2}$, in large-scale "weather" patterns are seen. (b) Similar to (a), for March 1995. Comparing with February, it can be seen that some non-radiative "weather cells" (indicated by shading) persist for at least a month and some (indicated by stippling) are more ephemeral. The largest features cover 10 to 20 cells, each computed independently.

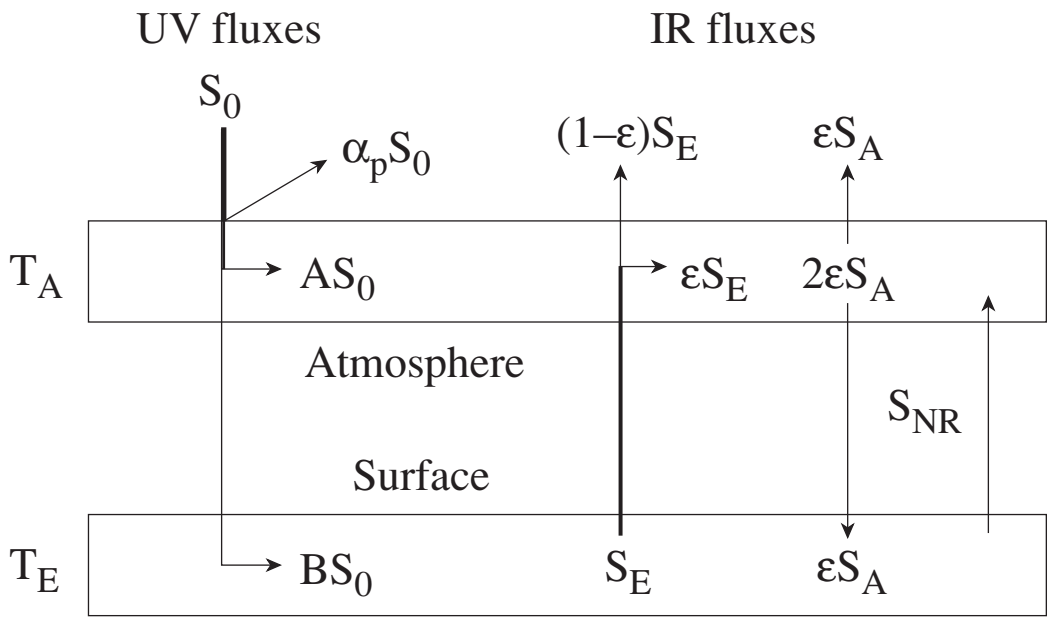

UV fluxes                          IR fluxes

$S_0$                    $(1-\varepsilon)S_E$        $\varepsilon S_A$

$\alpha_p S_0$

$T_A$          $AS_0$            $\varepsilon S_E$    $2\varepsilon S_A$

Atmosphere

$S_{NR}$

Surface

$T_E$          $BS_0$            $S_E$            $\varepsilon S_A$

E12086

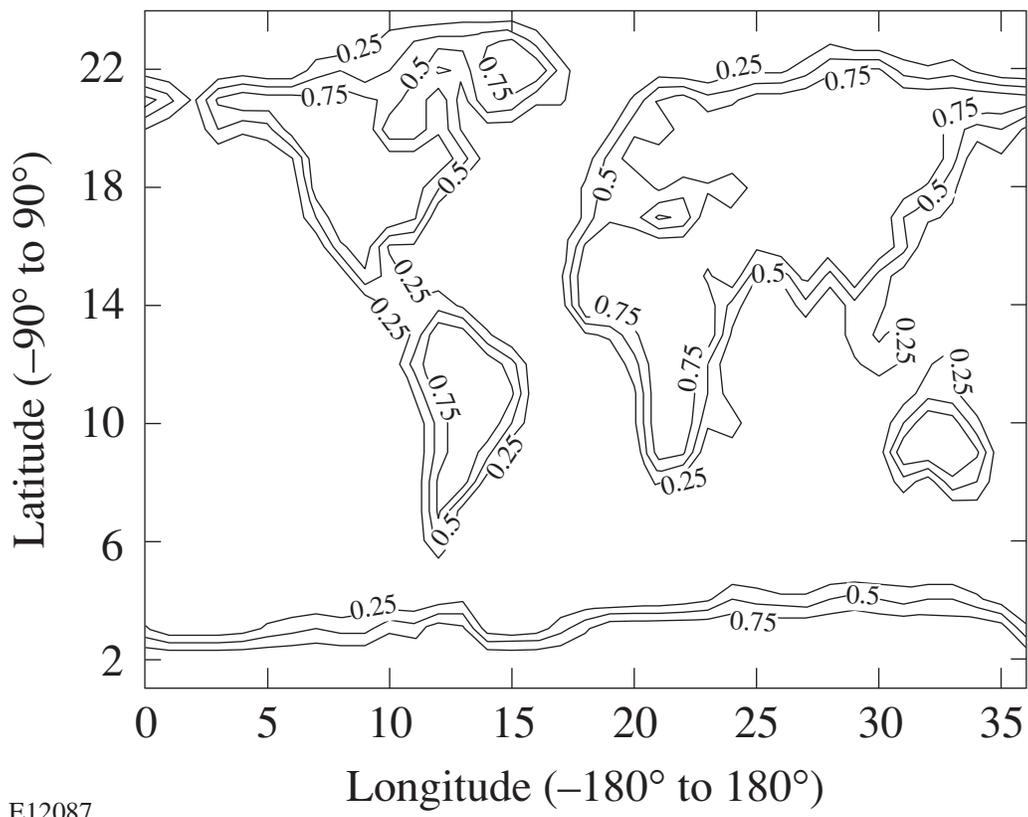

E12087

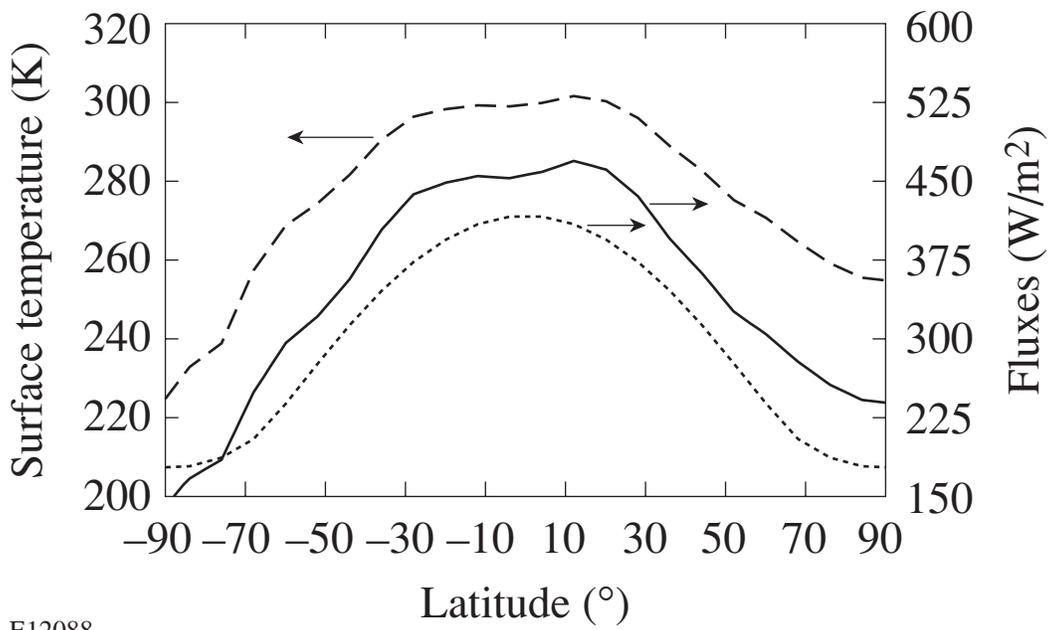

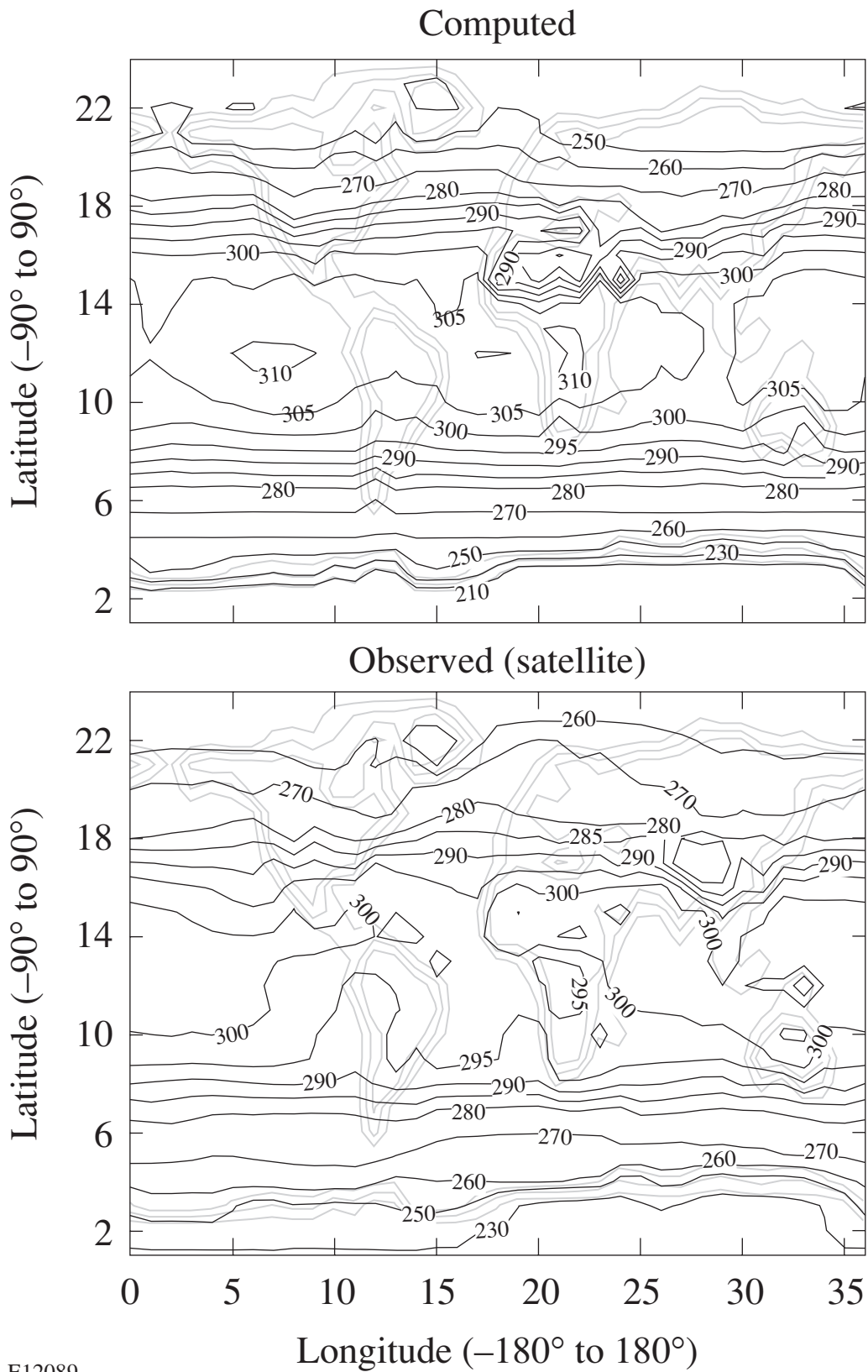

E12089

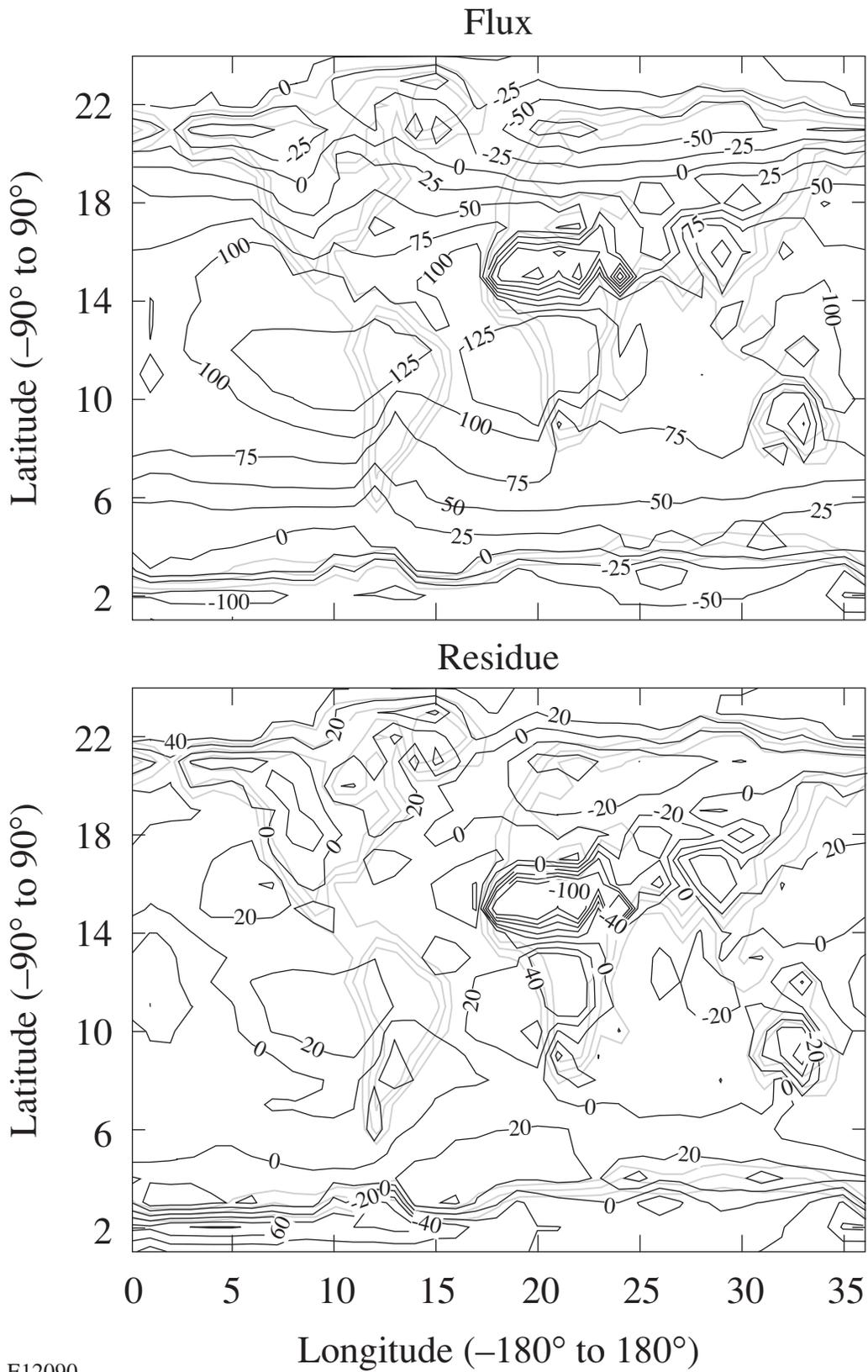

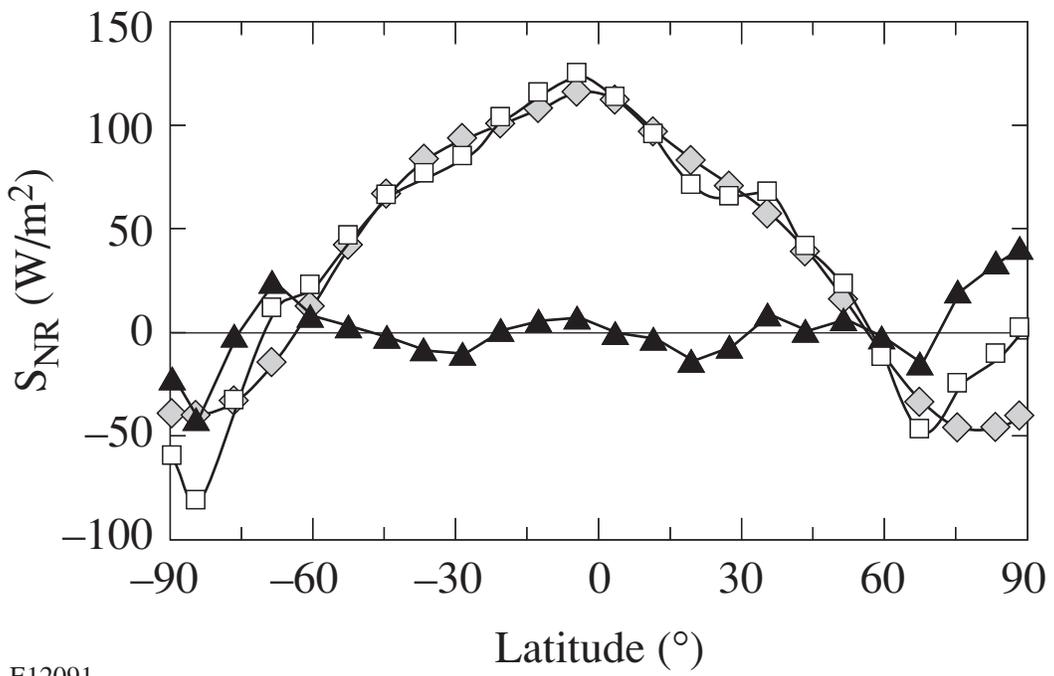

E12091

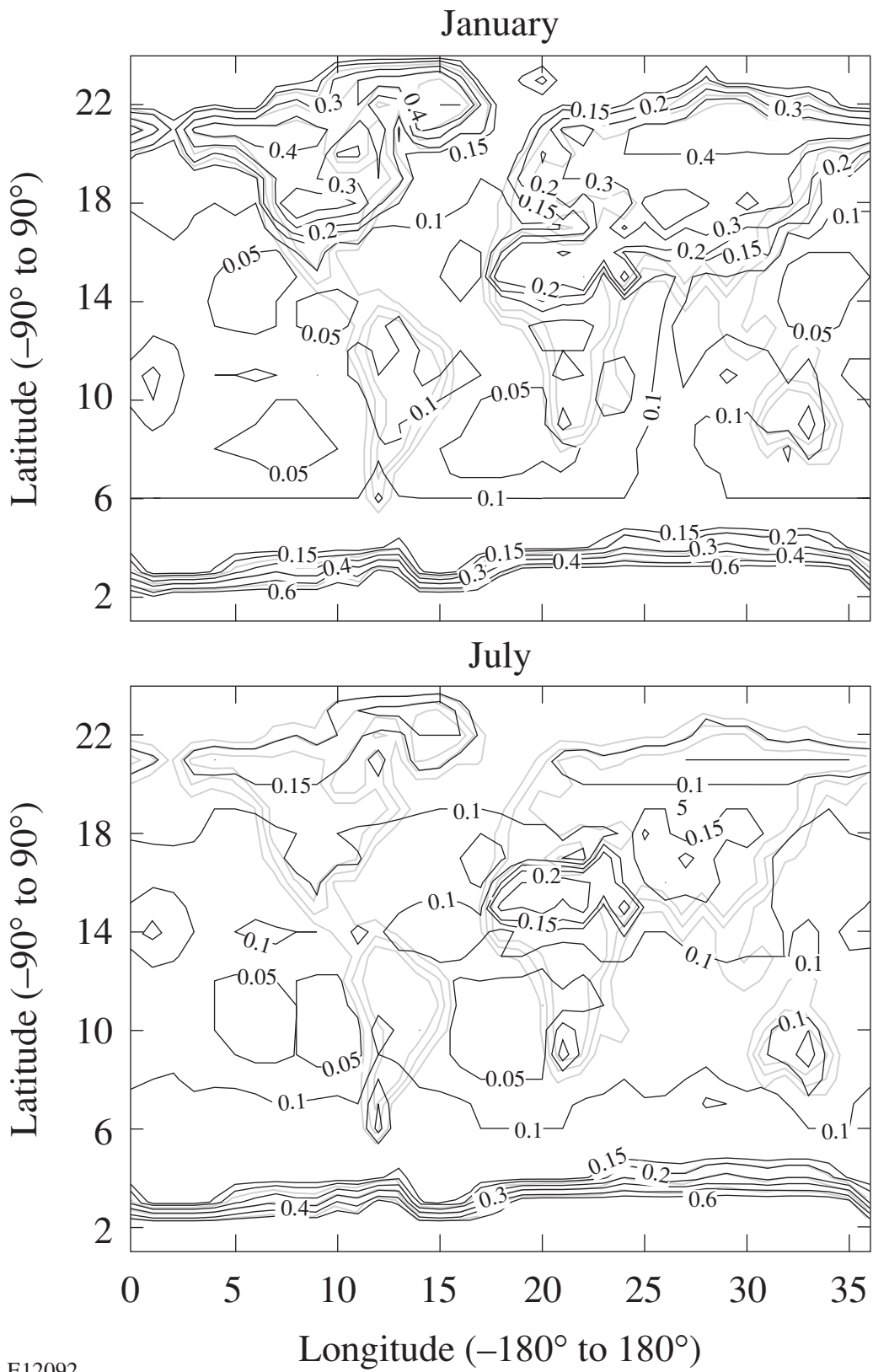

E12092

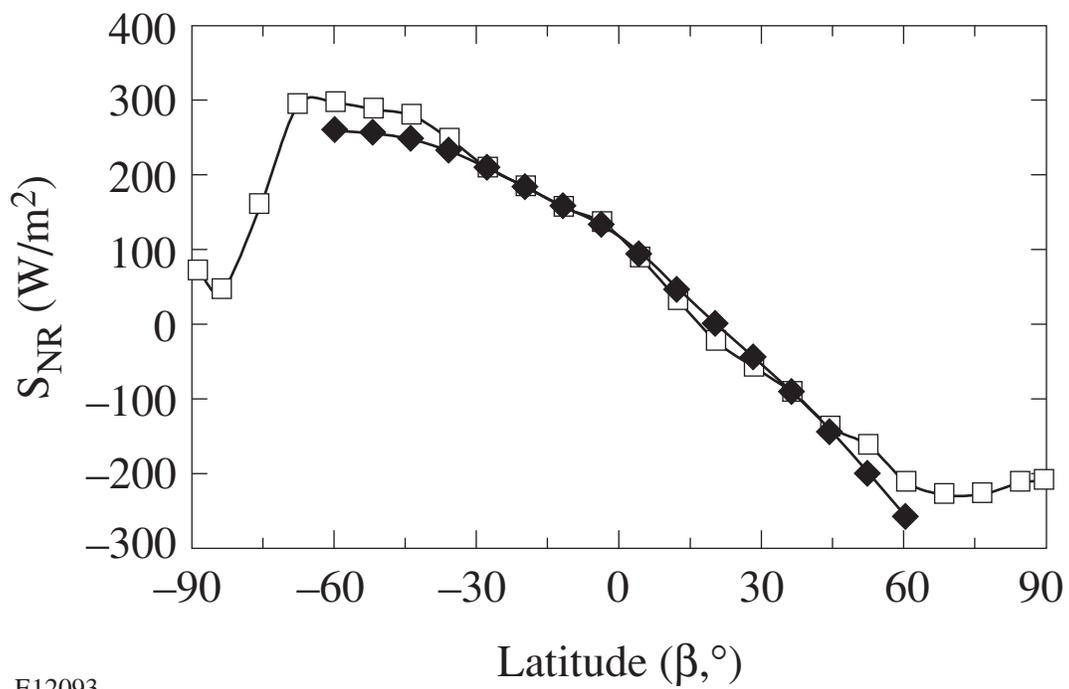

E12093

## February 1995

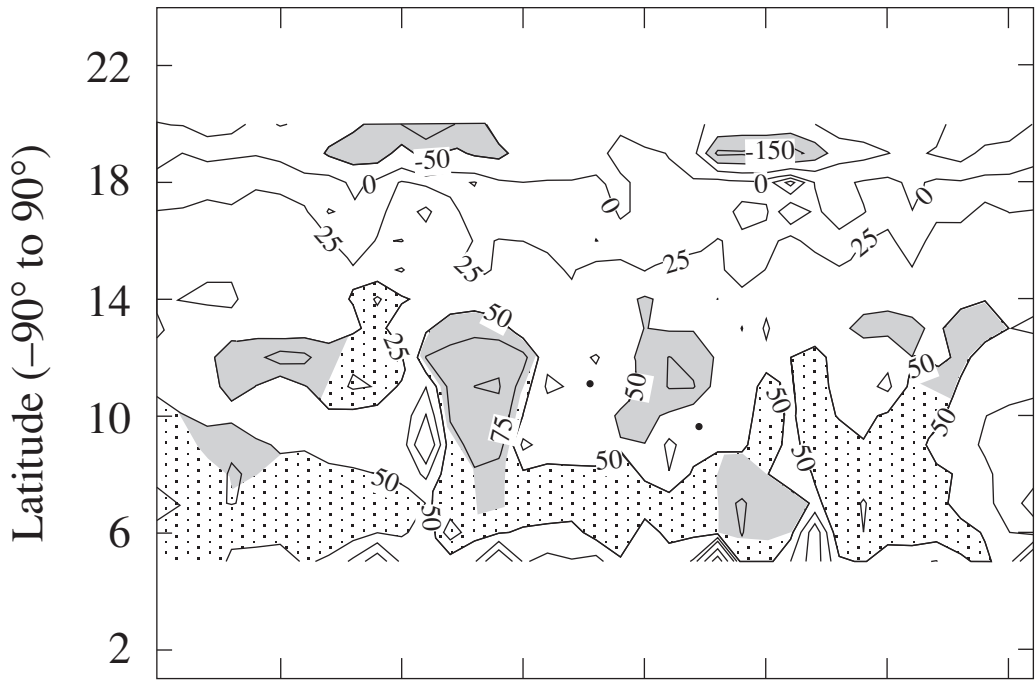

## March 1995

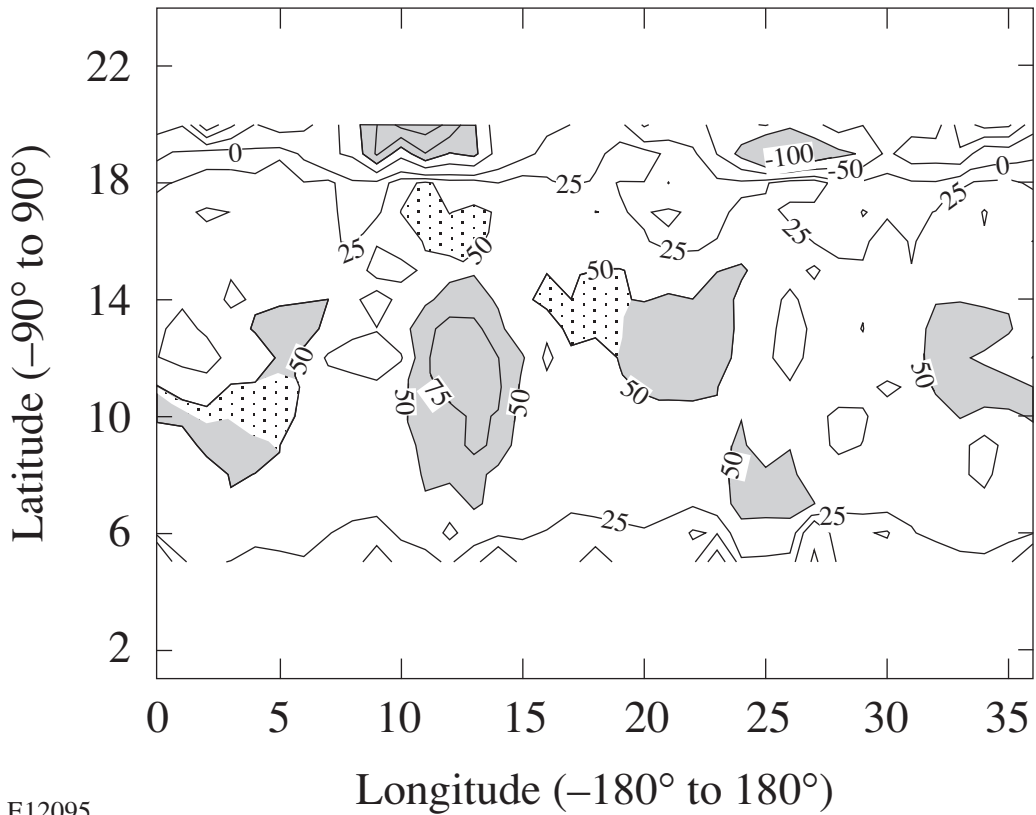

E12095

Longitude (−180° to 180°)